\documentclass[letter]{aa}
\usepackage[varg]{txfonts}
\usepackage{graphicx}
\usepackage{natbib}
\usepackage[figuresright]{rotating}
\usepackage{hyperref}
\usepackage[]{breakurl}

\bibpunct{(}{)}{;}{a}{}{,} 
\def\teff{$T\rm_{eff}$}
\def\kms{$\mathrm{km\, s^{-1}}$}

\newcommand{\Teff}{\ensuremath{T_\mathrm{eff}}}

\newcommand{\logg}{\ensuremath{\log g}}

\newcommand{\draftflag}{false}

\newcommand{\beq}{\begin{equation}}
\newcommand{\eeq}{\end{equation}}

\begin{document}
\idline{15}

\title{TOPoS: III. An ultra iron-poor multiple CEMP system\thanks{Based on 
observations made with ESO Telescopes at the La Silla Paranal Observatory under programme ID 094.D-0488 and 096.D-0616.}
}

\author{
E.~Caffau  \inst{1} \and
P.~Bonifacio   \inst{1} \and
M.~Spite       \inst{1} \and
F.~Spite       \inst{1} \and
L.~Monaco      \inst{2} \and
L.~Sbordone    \inst{3} \and
P.~Fran\c cois \inst{1,4} \and
A.~J.~Gallagher  \inst{1} \and
B.~Plez        \inst{5} \and
S.~Zaggia      \inst{6} \and
H.-G.~Ludwig   \inst{7,1} \and
R.~Cayrel      \inst{1} \and
A.~Koch        \inst{8} \and
M.~Steffen     \inst{9,1}\and  
S.~Salvadori   \inst{1} \and
R.~Klessen     \inst{10,11} \and
S.~Glover      \inst{10} \and
N.~Christlieb  \inst{7}
}

\institute{ 
GEPI, Observatoire de Paris, PSL Research University, CNRS, Univ. Paris Diderot, 
Sorbonne Paris Cit\'e, Place
Jules Janssen, 92190
Meudon, France
\and
Departamento de Ciencias Fisicas, Universidad Andres Bello, 220 Republica, Santiago, Chile
\and
European Southern Observatory, Casilla 19001, Santiago, Chile
\and
UPJV, Universit\'e de Picardie Jules Verne, 33 Rue St Leu, F-80080 Amiens
\and
Laboratoire Univers et Particules de Montpellier, LUPM, Universit\'e de Montpellier, 
CNRS, 34095 Montpellier cedex 5, France
\and
Istituto Nazionale di Astrofisica,
Osservatorio Astronomico di Padova Vicolo dell'Osservatorio 5, 35122 Padova, Italy
\and
Zentrum f\"ur Astronomie der Universit\"at Heidelberg, Landessternwarte, 
K\"onigstuhl 12, 69117 Heidelberg, Germany
\and
Phyics Department, Lancaster University, Lancaster LA1 4YB, UK
\and
Leibniz-Institut f\"ur Astrophysik Potsdam (AIP), An der Sternwarte 16, 14482 Potsdam, Germany
\and
Zentrum f\"ur Astronomie der Universit\"at Heidelberg,
Institut f\"ur Theoretische Astrophysik, Albert-Ueberle-Stra$\beta$e 2, 69120 Heidelberg, Germany
\and
Interdisziplin\"{a}res Zentrum f\"{u}r Wissenschaftliches Rechnen (IWR) der Universit\"{a}t Heidelberg
}
\authorrunning{189.D-0165}
\titlerunning{Stars Observed}
\date{Received ...; Accepted ...}

\abstract%
{}
{One of the primary objectives of the TOPoS survey is to search for the most metal-poor stars. 
Our search has led to the discovery of one of the most iron-poor objects known, SDSS\,J092912.32+023817.0. 
This object is a multiple system, in which two components are clearly detected in the spectrum. 
}
{We have analysed 16 high-resolution spectra obtained using the UVES spectrograph at the ESO 8.2\,m VLT
telescope to measure radial velocities and determine the chemical composition of the system. 
}
{Cross correlation of the spectra with a synthetic template
yields a double-peaked cross-correlation function (CCF) for eight
spectra, and in one case there is evidence for the presence of a third peak. 
Chemical analysis of the spectrum obtained by averaging all
the spectra for which the CCF showed a single peak found that
the iron abundance is [Fe/H]=--4.97. The system is 
also carbon enhanced with [C/Fe]=+3.91 ({\it A}(C) = 7.44).
From the permitted oxygen triplet we determined an upper limit
for oxygen of [O/Fe]$<+3.52$ such that C/O $>1.3$.
We are also able to provide more stringent
upper limits on the Sr and Ba abundances 
(${\rm [Sr/Fe]}<+0.70$, and ${\rm [Ba/Fe]}<+1.46$, respectively).
}
{}
\maketitle
\begin{figure}
\begin{center}
\resizebox{\hsize}{!}{\includegraphics[draft = \draftflag, clip=true]{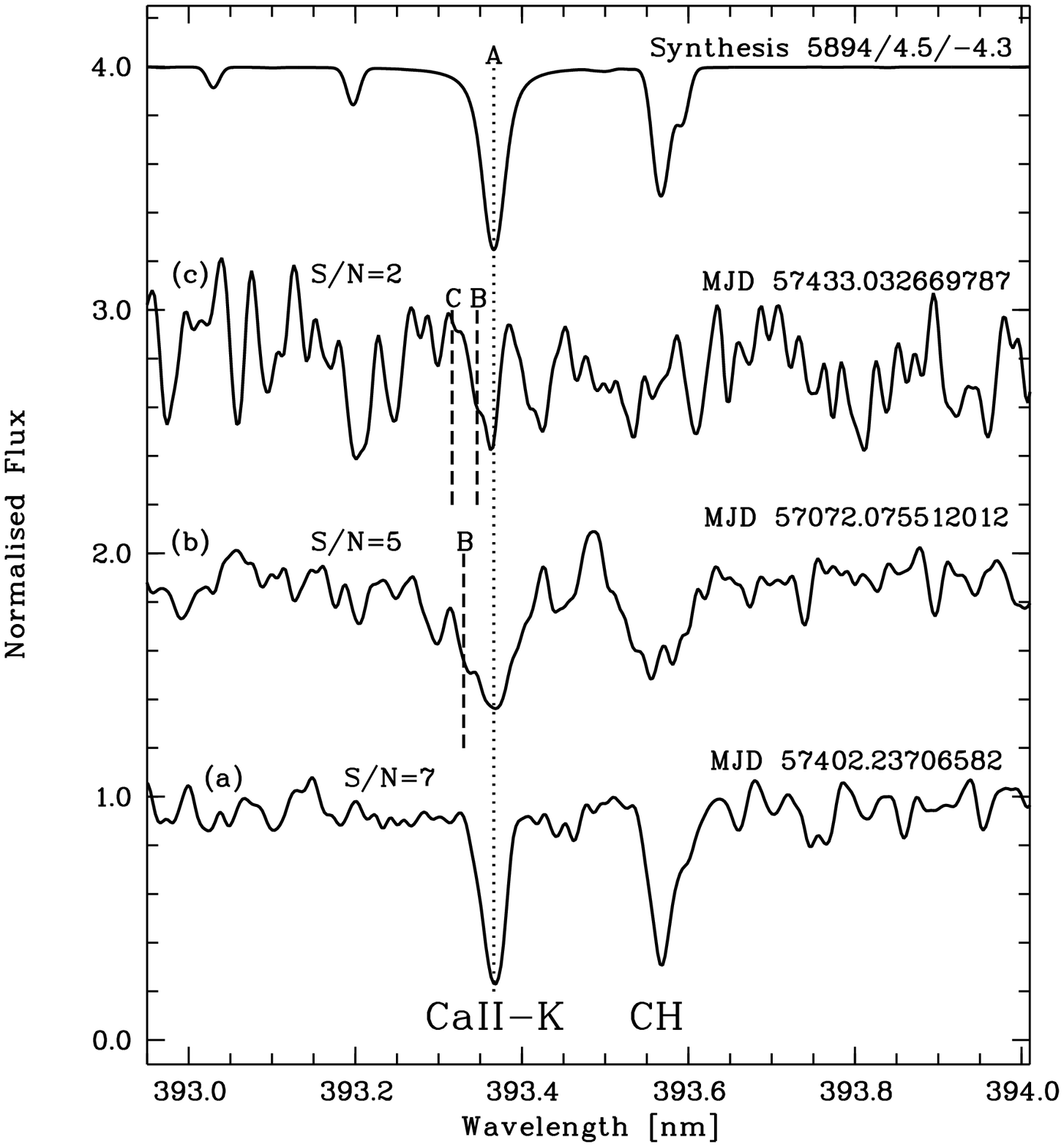}}
\resizebox{\hsize}{!}{\includegraphics[draft = \draftflag, clip=true]{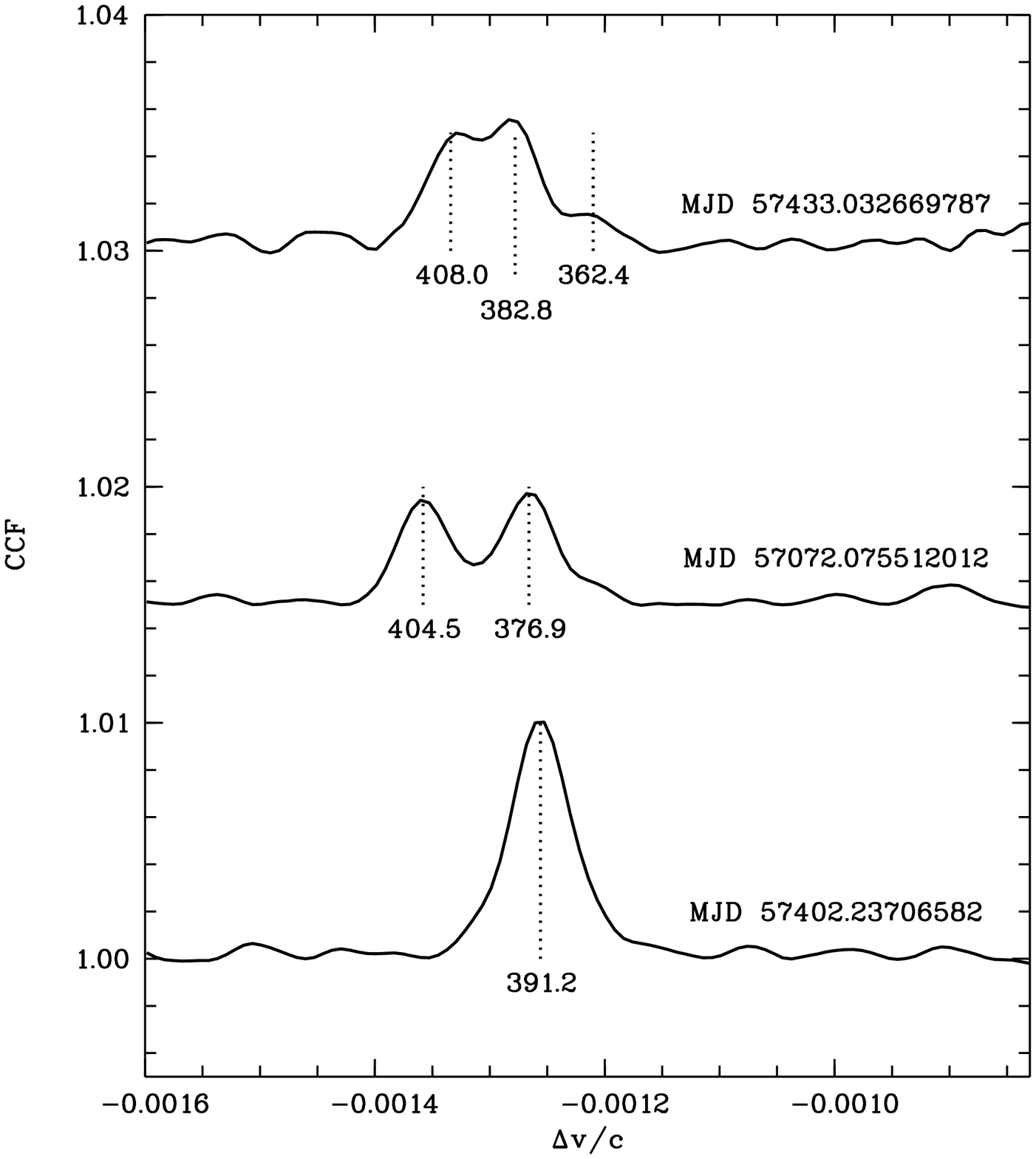}}
\end{center}
\caption[]{Upper panel: three representative
spectra of SDSS\,J0929+0238 in the range of \ion{Ca}{ii}-K (vertical dotted line) 
obtained on different dates and smoothed  by 10\,\kms.
The presence of several stars contributing to the spectra is evident from the clear
asymmetry of \ion{Ca}{ii}-K (highlighted by the vertical dashed lines) in the spectra 
corresponding to the multiple peak in the CCF.
Lower panel: the corresponding cross-correlation functions.
}
\label{plotccf}
\end{figure}

\section{Introduction}

Carbon enhanced metal-poor (CEMP) stars are characterised by a low iron content,
$[{\rm Fe/H}]\la-2.0$\,dex and an over-abundance in C, $[{\rm C/Fe}]>+1.0$\,dex \citep{bc05}.
CEMP stars very often show over-abundances of nitrogen and oxygen, 
sometimes show over-abundances of magnesium and sodium, and exhibit a wide distribution of heavy elements.
CEMP stars enriched in heavy elements, synthesised by both the slow (s-) 
and the rapid (r-) process, are referred to as CEMP-r/s stars; 
those enriched in heavy elements synthesised solely by the s-process are CEMP-s stars.
CEMP stars with a ``normal'' chemical pattern of the heavy elements 
(i.e. [Ba/Fe]$< +1.0$) are defined as CEMP-no stars.
\citet{spite13} pointed out that CEMP stars are divided into two groups 
according to their absolute carbon abundance
\citep[see also][]{masseron10}.
\citet{topos2} suggested that stars whose carbon has been accreted from 
a more evolved companion belong to the high-carbon band (with $A({\rm C})\approx 8.25$, \citealt{topos2}) 
while stars formed from Fe-poor gas clouds and showing no sign of late mass transfer belong to the low-carbon band
(with $A({\rm C})\approx 6.8$, \citealt{topos2}).
According to this picture, we expect all stars of the high-carbon band to be binary
\citep[see][]{lucatello05,Starkenburg,Hansen16,hansen16b}. In contrast, prior to 
the present paper, only seven CEMP-no stars 
were known to exhibit direct evidence of binarity.

We present the analysis of a binary, or perhaps trinary CEMP system,
SDSS\,J092912.32+023817.0 (henceforth referred to as SDSS\,J0929+0238), selected in the TOPoS project \citep{topos1}.
\begin{figure}
\begin{center}
\resizebox{\hsize}{!}{\includegraphics[clip=true]
{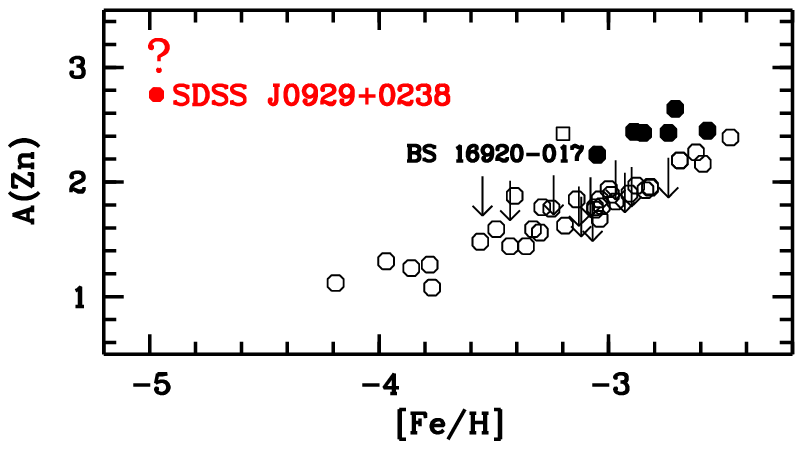}}
\end{center}
\caption[]{The zinc abundance as a function of metallicity.
The measurement for  BS\,16920-017 is from \citet{Honda}, that
of BD+44$^\circ$ 493 is from \citet{Roederer}.
the open circles are giants \citep{cayrel04}, filled
circles are dwarfs \citep{bonifacio09}, downward arrows are upper
limits for dwarfs \citep{bonifacio09}. The two red open
circles are CEMP-no stars.
The red full circle with a question mark is the Zn abundance
derived in the case of \logg =4.5 for SDSS\,J0929+0238.
}
\label{plotznfe}
\end{figure}

\section{Observations}

We observed SDSS\,J0929+0238 using the UV-Visual Echelle Spectrograph
\citep[UVES][]{dekker00} mounted at the Nasmyth platform B of the Unit\,2 
telescope (UT2) of the Very Large Telescope (VLT, Paranal observatory, Chile). 
Observations were conducted using the standard setting DIC2 437+760, 
which simultaneously covers  the wavelength ranges 373-499\,nm and 565-946\,nm, 
with its blue and red arms, respectively. Observations were collected in Service Mode  
during the period 8th February to 11th March, 2015 (program ID: 094.D-0488(A)) 
and 19th November, 2015 to 5th March, 2016 (096.D-0616(A)). A 1.6$^{\prime\prime}$ 
-wide slit was adopted for both arms, and the CCD was binned two by two.  
The spectral resolution corresponding to the adopted slit is R=28\,000 \relax and R=29\,000 \relax in 
the red and blue arms, respectively. A total of 16 observations with individual exposure times of 
3005\,s were taken. 
Spectra were reduced by the ESO staff
and retrieved from the ESO archive with the ESO Spectral Data Products Query Form.
Three
exposures not present in the archive were reduced using the UVES CLP-based
pipeline version  5.5.5.
Only the ``scired'' recipe was applied to the science frames, using the best 
reduced master calibrations associated to the frames by the archive query system.

\section{Analysis and Results}

The X-Shooter spectrum of SDSS\,J0929+0238 has been analysed in \citet{topos2}
with the stellar parameters \teff =5894\,K and \logg =3.7.
At that time, we pointed out a disagreement in the radial velocity
derived from the X-Shooter ($398\pm 10$\,\kms) and the SDSS ($467\pm 10$\,\kms) spectra.
We could not detect any iron lines and provide
an upper limit of ${\rm [Fe/H]<-3.81}$ from the X-Shooter spectrum. 
There are very few lines available to measure the radial velocity
from the individual exposures due to the low metallicity and relatively high temperature
of the system. 
Since the system is carbon-enhanced, the G-band is fairly prominent
and we found that it is suitable for radial velocity measurements.
We cross-correlated each exposure with a synthetic spectrum based on 
the results of 
\citet{topos2} (\Teff =5894\,K,  \logg =3.7, $A({\rm C})=7.7$, ${\rm [Fe/H]}=-4.5$).
In 8 out of 16 cases the cross correlation 
function (CCF) showed a clear double-peak and
in one case we have a tentative detection of a triple-peak.
For all 16 observations,
the radial velocities and the circumstances of the observation are listed in Table\,\ref{velhydra}.
When multiple peaks are present, we report the radial velocity corresponding to each peak.
When there is a single peak, the radial velocity
should be accurate to better than 1\,\kms, with the error dominated
by the uncertainty on the centring of the object on the slit
\citep[see e.g.][]{molaro}. However, when multiple peaks are
present, the error is dominated by the uncertainty
in the fitting of the multiple peaks, and, from repeated
measurements with slightly different fitting intervals, we estimate   
this to be as large as 2\,\kms.
However, we report all velocities with a precision
of 100\,ms$^{-1}$ to avoid rounding errors
in future period searches and orbital analyses. 

In the observation of 19th February 2015 (MJD=57072.075512012), the two peaks of the CCF are well separated (by 27.6\,\kms) 
and a secondary component of the \ion{Ca}{ii}-K line is 
visible in the  spectrum (see Fig.\,\ref{plotccf} for the observed spectrum and CCF denoted by (b)).
This object is therefore a double-lined spectroscopic (SB2) system, with evidence (CCF (c), 
Fig.\,\ref{plotccf}) that suggests it is possibly a 
triple-lined spectroscopic (SB3) system on the observation of 15th February 2015.
From the relative strengths of the cross-correlation peaks and the \ion{Ca}{ii}-K
line of the spectrum of 19th February 2015, we deduce that
the two main stars have to be very similar in temperature and luminosity.
This implies that they are both on the main sequence. If one of the two components
were on the sub-giant branch, its luminosity would be two magnitudes brighter, 
severely reducing the visibity of the companion's spectrum. 
We consider it very unlikely that both stars are on the sub-giant branch
since their masses should be identical to within 0.5\%.

In order to explore possible values for the \teff\ of the two components,
we compared the observed ${(g-z)}_0=0.511$ of the system to the 
theoretical colours obtained by combining the fluxes from pairs of stars
whose temperatures and luminosities lie along an isochrone of 14\,Gyr and
metallicity ${Z}=10^{-6}$ (Chieffi priv. comm.).
The combinations with ${ (g-z)}_0$ closest to the observed (all well within
the uncertainty, the uncertainty in the $g$ and $z$ magnitudes being of the order of 0.02) are:  
\begin{eqnarray*}
{\rm T}_{\rm A}=5950\,K~{\rm and}~{\rm T}_{\rm B}=5750\,K\ \Longrightarrow { (g-z)}_0=0.517,\\
{\rm T}_{\rm A}=5950\,K~{\rm and}~{\rm T}_{\rm B}=5800\,K\ \Longrightarrow { (g-z)}_0=0.508,\\
{\rm T}_{\rm A}=6000\,K~{\rm and}~{\rm T}_{\rm B}=5700\,K\ \Longrightarrow { (g-z)}_0=0.510, \\
{\rm T}_{\rm A}=6000\,K~{\rm and}~{\rm T}_{\rm B}=5750\,K\ \Longrightarrow { (g-z)}_0=0.503.
\end{eqnarray*}

We compared the combined synthetic spectrum in the range of the \ion{Ca}{ii}-K line to the spectrum
of 19th February 2015 that has 
the most separated double-peaked CCF (see CCF (b), Fig.\,\ref{plotccf}).
For a case where the primary star has a temperature $T_{\rm A}=6000$\,K, the secondary peak is less evident.
The shape of the line is best reproduced when the primary star has a temperature $T_{\rm A}\sim 5950$\,K
and the secondary is in the temperature range $5750<T_{\rm B}<5800$\,K.
For this small difference in \teff\ of less than 200\,K, the gravity of a main sequence star
changes very little ( $<0.05$\,dex), therefore we do not expect a significant amount of information from the gravity sensitive 
${(u-g)}$ colour that would disentangle the system.

We analysed the spectrum obtained by co-adding the seven spectra that have a single peaked CCF.
We assume that the stars in the system have been formed together and are not
the result of a tidal capture.
As stated above, the main stars in the system are at an evolutionary stage well before the first dredge-up takes place, 
so that neither the abundance of C or N have been changed.
In this picture we can safely assume that the stars share the same chemical composition.
Apart from the CH lines of the G-band, only six lines were clearly identified in the spectrum with 
a seventh {\it ad interim} detection.
To derive the chemical abundances, we fitted the line profiles \citep{abbo} with synthetic 
spectra computed by {\tt Turbospectrum} \citep{turbo}
based on MARCS model atmospheres \citep{G2008} used in \citet{topos2}
with an effective temperature derived from the colours of 5894\,K,
and two values of \logg: sub-giant star-like, \logg$= 3.7$, and main sequence star-like, \logg$=  4.5$.
We could not derive the micro-turbulence from the observed spectrum and so fixed it at 1.5\,\kms.
The abundances we derived are listed in Table\,\ref{abbo2} for both cases.
Since the two stars are very similar, the veiling correction, when the stars
appear at the same radial velocity (single peaked CCF), is very small.
We verified with synthetic spectra that if the two components have the same radial velocities
and their \teff\ are within 250\,K, the impact of veiling on the spectrum, when compared
to a single star synthesis, is well within our observational uncertainty.
In Table\,\ref{abbo2} we also report upper limits for O (\ion{O}{i} at 777.1\,nm), 
Sr (\ion{Sr}{ii} at 407.7\,nm) and Ba (\ion{Ba}{ii} at 455.4\,nm), based on a $3 \sigma$ 
deviation estimated by Cayrel's formula \citep{rcformula}.
For  Li, a signal-to-noise (S/N) ratio of approximately 70 \relax in the range  implies ${A({\rm Li})}<1.3$, at $3 \sigma$,
using the 3D-NLTE formula of \citet{sbordone10}; this put the stars in the Li-meltdown region.
The 3D corrections for the G-band would increase {\it A}(C) by approximately 0.05\,dex \citep[see][]{ajg16}.

At 4.6\,\kms\ to the blue (corresponding to 481.045\,nm) from the theoretical position of the \ion{Zn}{i} line at 481.0528\,nm
there is a feature with the shape of an absorption line. To our knowledge, in this metallicity regime,
there is no other possible identification for this line. 
We fitted the line profile assuming it is \ion{Zn}{i} and derived [Zn/H]=-1.86
(see Fig.\,\ref{plotznhydra}). The strength of the line with an equivalent width (EW) of $1.3$\,pm,
makes the line at the limit of detection. With a S/N of approximately 30 \relax in the range, the $3 \sigma$
EW detection, according to Cayrel's formula, is 1.0\,pm.
The weaker \ion{Zn}{i} line at 472.2153\,nm is not detected. However, the spectrum
is consistent with the presence of a line corresponding to the above Zn abundance
and shifted to the blue by 4.6\,\kms. 

\begin{table}
\caption{Abundances of the SDSS\,J0929+0238 system.}
\label{abbo2}
\renewcommand{\tabcolsep}{3pt}
\tabskip=0pt
\begin{center}
\begin{tabular}{lllll}
\hline\hline
\noalign{\smallskip}
Ion       & line    & Sun  &  \multicolumn{2}{c}{SDSS\,J0929+0238}  \\
          &  [nm]&  {\it A}(X)  &    [X/H]             &   [X/H]   \\
          &  &      &   \logg = 3.7        &  \logg =  4.5 \\
\hline \noalign{\smallskip}
\ion{Li}{i}& 607.7   &       & ${<}\phantom{-}1.24$       &  ${<}\phantom{-}1.27$  \\
\ion{C}{i} &G-band   & 8.50 & $\phantom{<}-0.81$         & $\phantom{<}-1.06$        \\
\ion{O}{i} &777.1   & 8.76 & $<-1.49$        & $<-1.21$       \\
\ion{Mg}{i}&383.8   & 7.54 & $\phantom{<}-4.63$         & $\phantom{<}-4.62$        \\
\ion{Ca}{ii}K&393.3 & 6.33 & $\phantom{<}-4.07$         & $\phantom{<}-4.20$        \\
\ion{Ca}{ii}T&866.2 & 6.33 & $\phantom{<}-4.73$         & $\phantom{<}-4.53$        \\
\ion{Fe}{i}&382.0,385.9   & 7.52 & $\phantom{<}-4.97\pm 0.11$ & $\phantom{<}-4.97\pm 0.11$\\
\ion{Zn}{i}&481.0   & 4.62 & $\phantom{<}-2.02$:         & $\phantom{<}-1.85$:        \\
\ion{Sr}{ii}&407.7  & 2.92 & $<-4.55$        & $<-4.27$       \\
\ion{Ba}{ii}&455.4  & 2.92 & $<-3.76$        & $<-3.51$       \\
\noalign{\smallskip}
\hline
\end{tabular}
\tablefoot{Solar abundances from \citet{abbosun} for Fe and C and from \citet{lodders09} for the other elements.
For all elements the uncertainty is between 0.15 and 0.2\, dex, for iron the quoted error
is the $\sigma$ of the two detected lines. A ``:'' denotes an uncertain measurement.
Li abundances are given as $A$(Li).}
\end{center}
\end{table}

\section{Discussion and Conclusions}

We have reported the discovery and analysis of a multiple system of CEMP stars.
The fact that cross-correlation on the G-band displays multiple peaks
on some dates implies that  all the components are CEMP stars.
In our view this makes it very unlikely that the C-enhancement is the result
of mass transfer from an AGB companion.
In the case of the CEMP-s double-lined binary
CS\,22964-161, \citet{Thompson} argued that this was the case
and suggested that the system was born as a hierarchical triple system
with a close binary that orbits a relatively low-mass AGB progenitor. 
If the presence of the third star is confirmed, this would make SDSS\,J0929+0238
even more complex, being the evolution of a hierarchical quadruple system.
In the following we do not elaborate on this scenario further 
and assume, as a working hypothesis, that the chemical composition of the stars
reflects that of the chemical composition of the cloud from which they have been 
formed.
The upper limits on Sr and Ba have been greatly improved with
respect to \citet{topos2}, however, they are still not stringent enough to allow
for classification of the system as a CEMP-no ([Ba/Fe]$< 1.0$, following  
\citealt{masseron10}) system. 
\citet{topos2} argued that stars on the low-carbon band are 
CEMP-no stars. \citet{Yoon16} arrived at a similar conclusion, albeit with a different definition of a CEMP star.
To confirm its CEMP-no nature, one would need a S/N$\approx 100$ in the region of the
\ion{Ba}{ii} resonance line. The 7h of observations analysed here only give a 
S/N = 35. This means that one would need to add approximately fifty more hours
of integration.  

There is a discrepancy between the Ca abundance derived from the
\ion{Ca}{ii}-K line and \ion{Ca}{ii} IR triplet.
Taking into account departures from NLTE worsens this discrepancy.
Looking at the NLTE corrections by \citet{stellina} for SDSS\, 
J102915+172927,whose parameters are similar, we see that the 
abundance from the K line should be corrected by --0.1\,dex and that
from the IR triplet, by --0.2\,dex. A straight average of the abundances
of the two lines implies [Ca/Fe]=+0.6 \relax in LTE and +0.45 \relax in NLTE; 
both are typical values for metal-poor stars.
 
The tentative detection of a  \ion{Zn}{i} line is surprising. The line 
is very weak, and the fact that it is shifted in wavelength casts some 
doubt on its identification. We have considered carbon-bearing molecules 
as responsible for this absorption and the only possible candidates are 
$\rm C_2$ lines that do not fall at the right wavelength, nor have the 
correct strength, even after considering the large C over-abundance implied 
by the G-band. Let us assume, for the sake of discussion, that the Zn 
abundance in the system is indeed as high as is implied by the absorption line
shown in Fig.\,\ref{plotznhydra}, [Zn/Fe]$\approx +3.0$. 
Figure\,\ref{plotznfe} depicts this star in the context of other EMP
stars from the First Stars project \citep{cayrel04,bonifacio09}.
The absolute zinc abundance of SDSS\,J0929+0238 is not different from what 
is found in more metal-rich stars, but it is very high for its exceptionally 
low Fe abundance. This makes its [Zn/Fe] ratio over two orders of magnitude 
larger than that seen elsewhere. Furthermore, SDSS\,J0929+0238 is a CEMP-no 
star, likely formed in a gas cloud polluted by primordial faint supernovae 
with mixing and fallback \citep[e.g.][]{topos2,salvadori15}.
These Pop~III stars only eject small amounts of zinc, implying that other 
sources must produce this heavy element while essentially not releasing iron. 
An increase in the [Zn/Fe] ratio at low [Fe/H] was noted by \citet{cayrel04} 
and attributed to $\alpha$-rich freeze-out processes. 
It is also significant that for ${\rm [Fe/H]}<-4.0$ there are no measurements 
of Zn abundance at all. To our knowledge Zn has only been measured in two 
other CEMP-no stars at [Fe/H]$\approx -4$: CS 22949--037 that has [Zn/Fe]=+0.7
\citep[][Fig.~\ref{plotznfe}]{Depagne} and BD+44$^\circ$ 493 \citep{Roederer}, 
[Zn/Fe]=-0.1. This suggests that there may be a real scatter in Zn abundances 
towards lower [Fe/H] as discussed in \citet{bonifacio09}. At higher [Fe/H], 
\citet{Honda} measured [Zn/Fe]=+1.0 \relax in BS\,16920-017 (Fig.~3) and 
\citet{Ivans} found almost the same value in CS\,22966-043 and G\,4-36, both 
at [Fe/H]$\approx -2.0$. Is SDSS\,J0929+0238 an extreme case of these Zn-rich stars?
Different sources have been suggested to produce high [Zn/Fe] ratios: bright and 
energetic ``hypernovae''\citep{Tominaga}, neutrino-driven winds in supernova 
progenitors \citep{Heger} and
low metallicity rotating stars  
(A. Chieffi, priv comm.). 
Yet none of the above possibilities can simultaneously 
produce such a high [Zn/Fe] ratio and low [Fe/H] value.

Only six CEMP-no stars are known to be binaries, including the prototype of the class, 
CS\,22957-027 \citep{Preston,Starkenburg,Hansen16}.
SDSS\,J0929+0238 is by far the most iron-poor among them and the only one that is suspected to be a triple system.
\citet{Latham} and \citet{carney03} have studied the fraction of binaries among Pop II stars based on data 
that spanned almost 20 years, and reached the conclusion that the 
binary fraction is not different from that found in Pop I stars \citep{Mayor}, with no
noticeable difference between dwarfs and giants. 
A similar conclusion was reached by \citet{Hansen16} for radial-velocity
measurements spanning eight years, targeting CEMP-no stars.

Our radial velocity data shows that the system is unmistakably a
binary or perhaps even a trinary system. 
One difficulty is that since the peaks in the CCFs are almost of equal
strength, it is ambiguous to assign a given velocity to the primary
or to the secondary. 
We have tried several methods to detect a period in the data:
the Lomb-Scargle periodogram \citep{lomb,scargle}, the Lafler-Kinman method \citep{LK} and
direct fitting of all the orbital parameters including period 
\citet{wichmann}. We also attempted direct fitting of the orbit 
with a Monte Carlo Markov Chain method \citep{koch2014}. This exercise led to several
plausible orbits with periods that range from 3 to 482 days.
Clearly, the data is too sparse to
detect a period or to attempt to derive an orbit. 
Continued monitoring
of this system is extremely important as the data will allow
us to constrain the masses of the components and determine whether this is a binary (most likely) or trinary system. 
This will convey important information about the formation of multiple systems 
of low-mass stars in the early Galaxy. 


\begin{acknowledgements}
The project is supported by fondation MERAC.
We are grateful to A. Chieffi for providing us with insight
into nucleosynthesis at low metallicity.
We acknowledge financial support from CNRS -INSU
Programme National de Cosmologie et Galaxies and
Programme National de Physique Stellaire.
This research has made use of the services of the ESO Science Archive Facility.
RSK acknowledges support from the European Research Council via the 
ERC Advanced Grant "STARLIGHT: Formation of the First Stars" (project number 339177).
HGL, RSK, SG and NC acknowledge financial support from Sonderforschungsbereich SFB 881 
"The Milky Way System" (subprojects A4, B1, B2 and B8) of the German Research Foundation (DFG).
SS is supported by a Marie Curie Fellowship, project 700907.
\end{acknowledgements}

\bibliographystyle{aa}

\appendix

\section{Additional data}

In Table\,\ref{velhydra} we report our measurements of the radial velocities.
To verify the robustness of the synthetic spectrum we used as a mask
for the cross-correlations, we used it to measure the radial velocity 
from a SOPHIE \citep{Bouchy,Perru08,Perru11} spectrum of HD\,201891. This star has similar stellar 
parameters (\teff=5850\,K and \logg=4.40, \citealt{mishenina13}) and its 
${\rm [Fe/H]}=-0.96$ makes its G-band of comparable strength to that shown in
the spectrum of SDSS\,J0929+0238.
We degraded the SOPHIE spectrum at the resolution of our UVES spectrum.
From the cross-correlation we derived a radial velocity of  0.4\,\kms smaller than the
result provided by the SOPHIE pipeline.
This tiny difference can be explained by the fact that in the range of the G-band
there are also metallic lines that we expect to be present in the spectrum of HD\,201891
but not in our mask. We consider the mask appropriate for providing radial velocities
to better than 1\,\kms.

\begin{table*}[!ht]
\sidecaption
\caption{Radial velocities derived from the G-band for SDSS\,J0929+0238. }
\label{velhydra}
\begin{center}
\begin{tabular}{llllr}
\hline\hline
\noalign{\smallskip}
Date   & UT               &   MJD                      & Vrad                           & BARYCOR   \\
       &                  &   (JD--2400000.5)          & [\kms]                         & [\kms]    \\
\noalign{\smallskip}                           
\hline                                         
\noalign{\smallskip}                           
2015-01-25&07:33:37.533   &   57047.315017751          &   390.56                       &$ 9.528105  $ \\
2015-02-19&01:48:44.237   &   57072.075512012          &   404.48/376.90                &$ -2.633427 $ \\
2015-11-20&07:03:07.093   &   57346.293832105          &   392.20                       &$ 29.748689 $ \\
2016-01-15&04:42:58.393   &   57402.196509185          &   391.35                       &$ 14.807685 $ \\
2016-01-15&05:41:22.486   &   57402.23706582           &   391.23                       &$ 14.688431 $ \\
2016-01-15&06:32:46.452   &   57402.27275987           &   391.72                       &$ 14.578024 $ \\
2016-01-15&03:51:40.927   &   57402.16089037           &   391.74                       &$ 14.902034 $ \\
2016-01-15&07:25:37.326   &   57402.309459796          &   391.90                       &$ 14.464483 $ \\
2016-02-14&03:06:44.969   &   57432.129687148          &   398.95/384.56                &$ -0.069693 $ \\
2016-02-14&02:15:09.096   &   57432.09385528           &   399.66/385.27                &$ 0.032276  $ \\
2016-02-15&00:47:02.669   &   57433.032669787          &   407.97/382.79/362.41         &$ -0.34363  $ \\
2016-03-05&00:50:42.637   &   57452.035215711          &   398.40/383.71                &$ -9.922114 $ \\
2016-03-05&01:43:58.919   &   57452.072209719          &   398.89/385.70                &$ -10.024746$ \\
2016-03-06&00:56:38.000   &   57453.03944384           &   388.91/374.22                &$-10.416031 $ \\
2016-03-06&01:50:02.879   &   57453.076422215          &   388.61/373.62                &$-10.520604 $ \\
2016-03-06&02:43:48.203   &   57453.113752352          &   387.71/373.02                &$-10.63447  $  \\
\noalign{\smallskip}
\hline
\end{tabular}
\end{center}
\tablefoot{When multiple peaks are detected in the CCF, the velocities reported in column 4 are ordered by the largest to the smallest.
We have not attempted to associate each velocity to one of the stellar components in the system.
The stars have very similar luminosities, therefore it is not easy to distinguish between them.
As such, to avoid giving misleading information this will be addressed when further data is secured.
The uncertainty is of the order of 1\,\kms\ for the single-peaked CCFs. 
Conservatively, we think it can be larger (within 2\,\kms ) due to the uncertainty in the fitting of multiple peaks.
}
\end{table*}

\begin{figure}[ht]
\begin{center}
\resizebox{\hsize}{!}{\includegraphics[clip=true]
{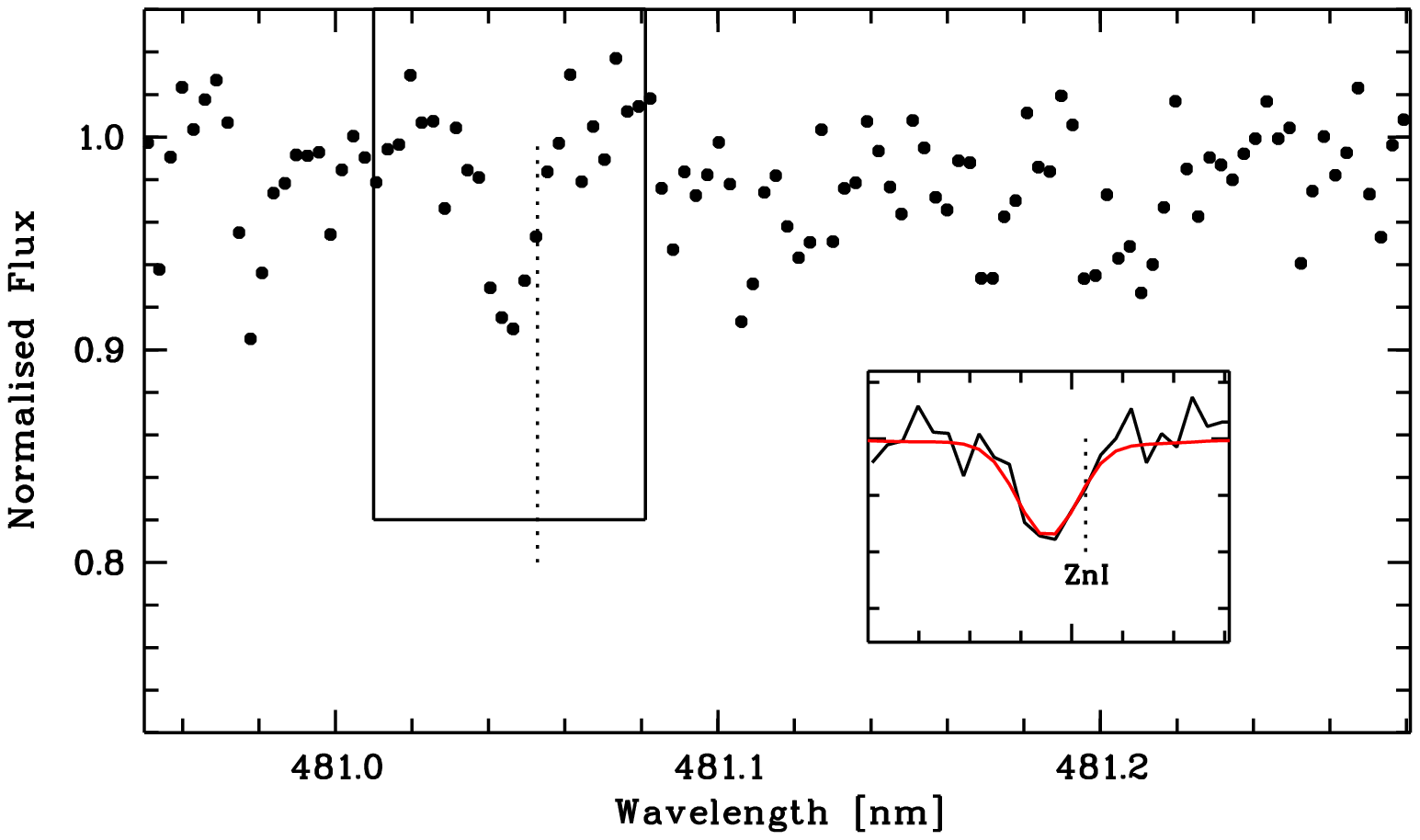}}
\end{center}
\caption[]{The spectrum of SDSS\,J0929+0238 \relax in the range of the \ion{Zn}{i} line (solid black).
The best fit (solid red) is shown in the inset for the case where \logg =4.5 and assuming the feature is Zn.
The vertical black dotted-line shows the laboratory position of the \ion{Zn}{i} 481.0528\,nm line.
}
\label{plotznhydra}
\end{figure}

In Fig.\,\ref{plotznhydra} we show the spectral range covering the \ion{Zn}{i} 481.0528\,nm line.
A zoom with the best fit is presented in the inset.
Formally the line is detected at $3.9 \sigma$, but Cayrel's formula does not take
into account the uncertainty in the continuum.
In practice, we consider the line to be detected at $3 \sigma$ and the shift
in wavelength makes the identification uncertain.
Using the TAPAS service \citep{tapas} we verified that there
are no telluric absorption lines in the range.

\end{document}